\documentclass[11pt]{article}
\usepackage{latexsym}
\usepackage{graphicx}
\usepackage{multirow}
\usepackage{amsmath}
\usepackage{ulem}
\usepackage{setspace}
\usepackage{subfigure}
\usepackage{color}
\usepackage{soul}
\begin{document}
\begin{center}\Large{Analytical computation of the Mercury perihelion precession via the relativistic gravitational law}
\end{center}
\begin{center}
\renewcommand\thefootnote{\fnsymbol{footnote}}
A.S. Fokas$^{1,2,*}$ and C.G. Vayenas$^{3,4,}$\footnote{E-mail: T.Fokas@damtp.cam.ac.uk; cgvayenas@upatras.gr}
\end{center}
\begin{center}
\textit{$^1${Department of Applied Mathematics and Theoretical Physics, University of Cambridge, Cambridge, CB3 0WA, UK}\\$^{2}${Department of Electrical Engineering, University of Southern California, Los Angeles, California, 90089-2560,USA}\\ $^3$LCEP, 1 Caratheodory St., University of Patras, Patras GR 26500, Greece\\$^4${Division of Natural Sciences, Academy of Athens, 28 Panepistimiou Ave., \\ GR-10679 Athens, Greece}} 
\end{center}

\begin{abstract}
{Let $r(\varphi)$ denote the orbit of Mercury. We compare the formulae obtained via general relativity for $r(\varphi)$ and for the corresponding perihelion precession angle $\Delta \varphi$, with the formulae obtained via the relativistic gravitational law, $F=GMm\gamma^6/r^2$. The latter law can be derived from Newton's gravitational law by employing the gravitational rather than the rest masses of the Sun and Mercury. Remarkably, it is found that the two expressions for the dominant part of $r(\varphi)$ are identical and that the two expressions for $\Delta\varphi$ are also identical.}
\end{abstract}
\vspace{0.5cm}
\vspace{0.5cm}

\section{Introduction}
The experimental verification of the formulae for the perihelion precession of Mercury and for the bending angle of light passing near massive bodies such as our Sun, are the two most  spectacular confirmations of Einstein's theory of general relativity \cite{Einstein1916}. Furthermore, the efforts of Einstein in order to explain these two phenomena played an important role in the development of general relativity (GR) \cite{Linton04}. It is generally accepted that Newton's universal gravitational law, which cannot account for the observed perihelion advancement and which gives only half of the experimentally observed light bending angle, is not accurate enough under relativistic conditions \cite{Misner09,Das11}. 

However, when employing Newton's universal gravitational law under relativistic conditions, little attention has been paid to the fact that there exist differences between the rest, relativistic, inertial and gravitational masses, of the bodies involved. 

The inertial mass, $m_i$, and the gravitational mass, $m_g$, are equal according to the equivalence principle \cite{Roll1964}, but the ratio of the inertial mass to the rest mass and also the ratio of the relativistic mass, $\gamma m_o$, to the rest mass are unbounded as the particle velocity approaches the speed of light \cite{Einstein1905}. 

Consequently, the following important question arises when using Newton's universal gravitational law under conditions where the Lorentz factor $\gamma$ is significantly larger than unity: should one use the particle rest mass, $m_o$, the relativistic mass, $\gamma m_o$, or the gravitational mass, $m_g$, which equals the inertial mass, $m_i$? This question has been addressed recently in \cite{Vayenas12, Vayenas1106, Vayenas15, Vayenas16}, where it is argued that one should use the gravitational mass, $m_g$, rather than the rest mass, $m_o$, or the relativistic mass, $\gamma m_o$. Perhaps one of the reasons that this question was not discussed earlier is that the velocities of the planets of our planetary system are of the order $10^{-4}c$, thus the corresponding Lorentz factors are of the order of $1+10^{-8}$ and therefore the differences between rest, relativistic, and gravitational masses are very small, which is also the case for the magnitude of the perihelion precession value per orbit and of the light bending angle caused by our Sun. However,this is not the case for many other astronomical systems involving heavier stars \cite{Bertone10,Das09}.

The inertial mass, $m_i$, and hence according to the equivalence principle, the gravitational mass $m_g$, is related with the rest mass, $m_o$, via the equation
\begin{equation}
\label{eq1}
m_g=m_i=\gamma^3m_o,\quad \gamma=(1-\texttt{v}^2/c^2)^{-1/2}.
\end{equation}

This result for linear motion was first obtained in Einstein's pioneering special relativity paper \cite{Einstein1905,French68,Freund08}; it has been argued recently \cite{Vayenas12,Vayenas1106} that this result remains valid for an arbitrary particle motion, including circular motion. 

Consequently, Newton's universal gravitational law for the gravitational force between two particles, at a distance $r$, of rest masses $m_{o,1},m_{o,2}$ and moving with a velocity $\texttt{v}$ relative to each other, is given by
\begin{equation}
\label{eq2}
F=\frac{Gm_{g,1}m_{g,2}}{r^2}=\frac{Gm_{o,1}m_{o,2}\gamma^6}{r^2}\;.
\end{equation}

If $m_{o,1}=m_{o,2}=m_o$, the above equation reduces to 
\begin{equation}
\label{eq3}
F=\frac{Gm^2_o\gamma^6}{r^2}.
\end{equation}

By employing equation (\ref{eq3}) it is possible to predict the existence of gravitationally confined circular rotational states with many of the properties of hadrons \cite{Vayenas12, Vayenas1106, Vayenas15}, or bosons \cite{Vayenas16}, and to show that the rest masses of quarks are very small, in the mass range of neutrinos \cite{Vayenas12,Vayenas1106}. Furthremore, using equation (\ref{eq3}) it can be shown that the ratio $F/F_o$ of the force, $F$, keeping two or three-particle systems in orbit, divided by the same force computed for $\gamma=1$, is $(m_{Pl}/m_o)^2$, which is exactly the value predicted by GR \cite{Vayenas12, Vayenas1106}.

\begin{figure}[t]
\vspace{1cm}
\includegraphics[width=0.25\textwidth]{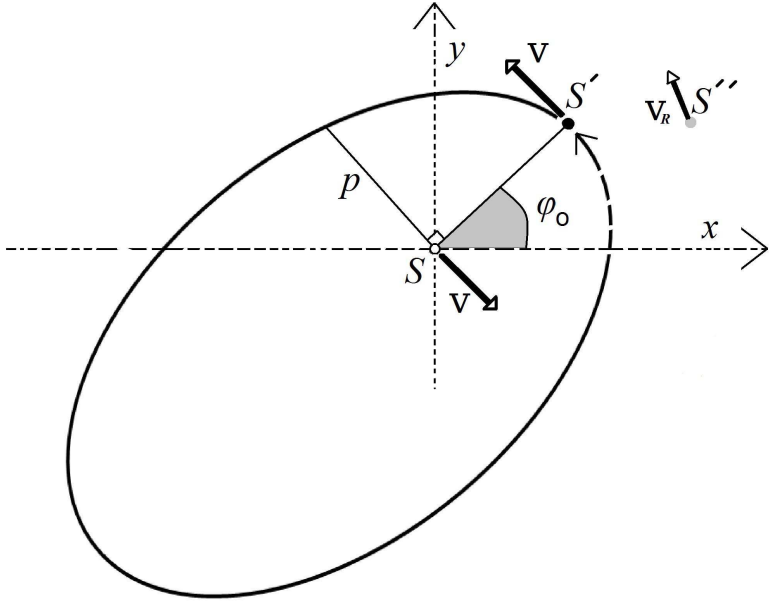}
\vspace{2cm}
\caption{Model geometry showing the coordinate system and the definitions of $\varphi_o$ and $p$.}
\label{fig:1}
\end{figure}

For the motion of a particle with rest mass $m_o$ rotating around a massive particle of rest mass $M_o$, equation (\ref{eq2}) remains valid. Indeed, consider for example the Sun (mass $M_o$) and Mercury (mass $m_o$) at the locations of two instantaneous inertial systems, $S$ and $S'$, (Fig. 1) moving with an instantaneous velocity $\texttt{v}$ relative to each other. In this case, the inertial and thus the gravitational inertial mass of Mercury is given by $\gamma^3m_o$ according to the observer on the Sun at $S$, but also the inertial mass of Sun is given by $\gamma^3M_o$ according to the observer on Mercury at $S'$.

If one uses rest masses instead of inertial masses then, according to Newtonian mechanics the particle of mass $m_o$ moves on an ellipse. Consider such an ellipse with an elliptical parameter $p$ (also known as the semi-latus rectum), with eccentricity $e$, and with a semi-major axis $a$; these parameters are related by the equation
\begin{equation}
\label{eq4}
p=(1-e^2)a.
\end{equation}

GR implies the following expression for the precession angle, $\Delta \varphi$, \cite{Vankov11,Misner73,Weinberg72,Wald84,Taylor2000,Magnan07}
\begin{equation}
\label{eq5}
\Delta\varphi=\frac{6\pi GM}{pc^2}, \quad p=\frac{J^2}{GM},
\end{equation}
where $J$ denotes angular momentum per unit mass.

Using the momentum-velocity equation of special relativity (SR), namely momentum equals $\gamma m_o\texttt{v}$, Lemmon and Mondragon examined in \cite{Lemmon10} the resulting corrections to Keplerian orbits and to the precession angle $\Delta \varphi$. Their work, which does not take into consideration equation (\ref{eq1}), yields one sixth of the value computed via GR (equation (\ref{eq5})). 

It is interesting to note that Silberstein already back in 1917 \cite{Silberstein17} attempted to account for the velocity dependence of the rotating mass $m$ by combining $m=\gamma m_o$ with the empirical force expression
\begin{equation}
\label{eq6}
F=\frac{GMm_o}{r^2}f(\gamma).
\end{equation}

He reported that he could get agreement with the GR result of equation (\ref{eq5}) provided that $f(\gamma)=\gamma^{n-1}$ and $n=6$. He noted that it was entirely unclear to him why this particular value of $n$ leads to exact agreement with GR \cite{Silberstein17}. Furthermore, Silberstein only analysed the relevant equations in the limit of $\texttt{v}/c\rightarrow 0$, where $\texttt{v}$ is the velocity of the rotating particle.

Here, for the sake of completness we first present a simple rederivation of the well known \cite{Das11,Vankov11,Misner73,Weinberg72,Wald84,Taylor2000,Magnan07} GR expression
\begin{equation}
\label{eq7}
\frac{d^2U}{d\varphi^2}+U=1+\left(\frac{3\varepsilon}{2}\right)U^2,
\end{equation}
where
\begin{equation}
\label{eq8}
U=\frac{J^2}{GMr}\quad,\quad J=\frac{L}{m}\;,\quad r_s=\frac{2GM}{c^2}\;,\quad \varepsilon=\frac{r_s}{p}=2\left(\frac{GM}{cJ}\right)^2.
\end{equation}
In equations (\ref{eq8}), $L$ is the angular momentum of the particle of mass $m$, $M$ is the mass of the heavier body, $G$ is the gravitational constant, and $J$ is the angular momentum per unit mass.

We then present the first main result of our paper: if a particle of mass $m$ moves around a heavier mass $M$ under the influence of the relativistic gravitational law (\ref{eq2}), then $\widetilde{U}$ satisfies the ordinary differential equation
\begin{equation}
\label{eq9}
\frac{d^2\widetilde{U}}{d\varphi^2}+\widetilde{U}=\left\{1-\frac{\widetilde{\varepsilon}}{2}\left[\widetilde{U}^2+\left(\frac{d\widetilde{U}}{d\varphi}\right)^2\right]\right\}^{-3},
\end{equation}
where
\begin{equation*}
\widetilde{U}=\frac{\widetilde{J}^2}{GM\widetilde{r}}\quad,\quad \widetilde{J}=\frac{\widetilde{L}}{m_o}\;,\quad \widetilde{\varepsilon}=\frac{r_s}{\widetilde{p}}=2\left(\frac{GM}{c\widetilde{J}}\right)^2,\quad \widetilde{p}=\frac{\widetilde{J}^2}{GM}
\end{equation*}
and $\; \widetilde{}\;$  denotes quantities computed via equations (\ref{eq1}) to (\ref{eq3}).

It turns out that for small $\varepsilon$ the dominant parts of the solutions of equations (\ref{eq7}) and (\ref{eq9}) are both given by the expression
\begin{equation}
\label{eq10}
U_D=\left(\frac{3\varepsilon}{2}\right) e\varphi\sin\varphi,
\end{equation}
and thus, the precession angle per revolution, $\Delta\varphi$, is given in both cases by
\begin{equation}
\label{eq11}
\Delta\varphi=2\pi \left(\frac{3\varepsilon}{2}\right),
\end{equation}
which using the definition of $\varepsilon$ in the third of equations (\ref{eq8}) becomes the well known formulae (\ref{eq5}).

It is important to note that one may consider a third instantaneous inertial system, $S''$, with an observer, located on earth and with a velocity $\texttt{v}_R$ with respect to one of the two moving bodies. This second velocity $\texttt{v}_R$, and the corresponding Lorentz factor, $\gamma_R$, defines the ratio
\begin{equation}
\label{eq12}
m_r/m_o=\gamma_R,
\end{equation}
of the relativistic and rest masses of Mercury as observed from $S''$, i.e. from the Earth. Since $m_o$ is not amenable to direct observation, and also since $\Delta \varphi$ should be independent of $S''$, we may set $\gamma_R=1$.

\section{General Relativistic Treatment Revisited}
Let $(r,\varphi)$ be cylindrical coordinates in a plane containing the center of the sun, and let $\tau$, $E$, and $L$ denote the proper time, energy, and angular momentum respectively.

The basic equations of GR are conservation of energy and of angular momentum, as well as the Schwarzschild metric equation:
\begin{equation}
\label{eq13}
\frac{E}{mc^2}=\left(1-\frac{r_s}{r}\right)\frac{dt}{d\tau},
\end{equation}

\begin{equation}
\label{eq14}
\frac{L}{m}=r^2\frac{d\varphi}{d\tau},
\end{equation}

\begin{equation}
\label{eq15}
c^2d\tau^2=c^2\left(1-\frac{r_s}{r}\right)dt^2-\left(1-\frac{r_s}{r}\right)^{-1}dr^2-r^2d\varphi^2,
\end{equation}

\begin{equation}
\label{eq16}
r_s=\frac{2GM}{c^2},
\end{equation}
where $r_s$ is the Schwarzschild radius of the sun and $M$ is the mass of the sun.

Solving equations (\ref{eq13}) and (\ref{eq14}) for $dt$ and $d\tau$ in terms of $d\varphi$, substituting the resulting expressions into (\ref{eq15}), and then solving the resulting equation for $(dr/d\varphi)^2$ we find
\begin{equation}
\label{eq17}
\left(\frac{1}{r^2}\frac{dr}{d\varphi}\right)^2=\frac{E^2}{L^2c^2} -\left(1-\frac{r_s}{r}\right)\left(\frac{m^2c^2}{L^2}+\frac{1}{r^2}\right).
\end{equation}

Making the substitution
\begin{equation*}
u=\frac{1}{r},
\end{equation*}
equation (\ref{eq17}) becomes
\begin{equation}
\label{eq18}
\left(\frac{du}{d\varphi}\right)^2=\frac{E^2}{L^2c^2}-\left(1-r_s u\right) \left(\frac{m^2c^2}{L^2}+u^2\right).
\end{equation}

Differentiating this equation with respect to $\varphi$ and introducing the dimensionless quantities defined in equations (\ref{eq8}) we find equation (\ref{eq7}).

\section{Motion Due to a Central Force}
Consider a single particle of relativistic mass $m_r$ and of gravitational mass $m_g$ moving under the influence of a central force of magnitude $f$. Then,
\begin{equation}
\label{eq19}
\frac{d}{dt}(m_r\underline{\dot{r}})=-f \underline{\hat{r}}, \quad \quad \hat{\underline{r}}=\frac{\underline{r}}{\left|\underline{r}\right|}.
\end{equation}

Incidentaly, in the particular case that $m_r=m_o\gamma$, using the definition of $\gamma$, i.e. the second of equations (\ref{eq1}), we find
\begin{equation}
\label{eq20}
\frac{d\gamma}{dt}=\frac{\gamma^3}{c^2}\texttt{v}\frac{d\texttt{v}}{dt},\quad \quad \texttt{v}=\left|\underline{\dot{r}}\right|.
\end{equation}

Hence,
\begin{equation}
\label{eq21}
\frac{d}{dt}(m_o\gamma \underline{\dot{r}})=m_o\gamma \underline{\ddot{r}}+\frac{m_o\gamma^3}{c^2}\left(\texttt{v}\frac{d\texttt{v}}{dt}\right)\underline{\dot{r}}.
\end{equation}
It is shown in the Appendix that
\begin{equation*}
\texttt{v}\frac{d\texttt{v}}{dt}=\underline{\dot{r}}\cdot \underline{\ddot{r}},
\end{equation*}
and then equation (\ref{eq21}) takes the familiar form found in most books, namely
\begin{equation}
\label{eq22}
\frac{d}{dt}(m_o\gamma \underline{\dot{r}})=m_o\gamma \underline{\ddot{r}}+\frac{m_o\gamma^3}{c^2}\left(\underline{\dot{r}}\cdot \underline{\ddot{r}}\right)\underline{\dot{r}}.
\end{equation}

Starting with (\ref{eq19}), it is straightforward to derive the following results:
\begin{enumerate}
	\item The angular momentum $\underline{L}$ is conserved, where $\underline{L}$ is defined by
	\begin{equation}
\label{eq23}
\underline{L}=\underline{r}\times m_r\underline{\dot{r}}.
\end{equation}
Furthermore, the magnitude of the angular momentun denoted by $L$ is given by
\begin{equation}
\label{eq24}
L=m_rr^2\frac{d\varphi}{dt},
\end{equation}
where $(r, \varphi)$ are cylindrical coordinates in the plane of motion which clearly is orthogonal to $L$.

\item $u(\varphi)$ satisfies the equation
\begin{equation}
\label{eq25}
\frac{d^2u}{d\varphi^2}+u=\frac{fm_r}{(Lu)^2},\quad \quad u=\frac{1}{r}.
\end{equation}

\item The speed of the particle $\texttt{v}$ satisfies the equation
\begin{equation}
\label{eq26}
\texttt{v}^2=\left(\frac{L}{m_r}\right)^2\left[u^2+\left(\frac{du}{d\varphi}\right)^2\right].
\end{equation}
\end{enumerate}

In order to derive the above results we begin by differentiating equation (\ref{eq23}) with respect to $t$:
\begin{equation}
\label{eq27}
\underline{\dot{L}}=\underline{\dot{r}}\times m_r\underline{\dot{r}}+\underline{r}\times\frac{d}{dt}(m_r\underline{\dot{r}})=\underline{r}\times(-f\underline{\hat{r}})=0.
\end{equation}
Thus, $\underline{L}$ is constant.

Equation (\ref{eq23}) implies that $\underline{r}$ is perpendicular to $\underline{L}$.

Introducting cylindrical coordinates in the plane perpendicular to $\underline{L}$, we find
\begin{equation}
\label{eq28}
\underline{r}=r\underline{\hat{r}},\quad \underline{\hat{r}}=\cos\varphi\underline{i}+\sin\varphi\underline{j},
\end{equation}
where $\underline{i}$ and $\underline{j}$ are unit vectors along the $x$ and $y$ axis.

Differentiating (\ref{eq28}) with respect to $t$ we find
\begin{equation}
\label{eq29}
\underline{\dot{r}}=(\dot{r}\cos\varphi-r\dot{\varphi}\sin\varphi)\underline{i}+(\dot{r}\sin\varphi+r\dot{\varphi}\cos\varphi)\underline{j}.
\end{equation}

Thus, the definition of $\underline{L}$ implies
\begin{equation}
\label{eq30}
\underline{L}=m_rr^2\dot{\varphi}\underline{\kappa},
\end{equation}
where $\underline{\kappa}$ is the unit vector perpendicular to $\underline{L}$.

Let $L$ denote the magnitude of $\underline{L}$. Equation (\ref{eq30}) implies
\begin{equation}
\label{eq31}
\dot{\varphi}=\frac{L}{m_rr^2}.
\end{equation}

Decomposing equation (\ref{eq19}) along the $x$ and $y$ axis we find the following equations:
\begin{equation}
\label{eq32}
\frac{d}{dt}(m_r\dot{x})=-f\cos\varphi, \quad \frac{d}{dt}(m_r\dot{y})=-f\sin\varphi.
\end{equation}
Using the indentity $r=1/u$ we find
\begin{equation}
\label{eq33}
m_r\dot{x}=m_r\frac{d}{dt}(r\cos{\varphi})=m_r\dot{\varphi}\frac{d}{d\varphi}(u^{-1}\cos\varphi).
\end{equation}
Replacing in this equation $\dot{\varphi}$ via equation (\ref{eq31}) and simplifying we find
\begin{equation}
\label{eq34}
m_r\dot{x}=-L\left(u\sin\varphi +\cos\varphi\frac{du}{d\varphi}\right).
\end{equation}

Thus, the first of equations (\ref{eq32}) yields
$$-L\dot{\varphi}\left(u\cos\varphi+\frac{du}{d\varphi}\sin\varphi-\sin\varphi\frac{du}{d\varphi}+\cos\varphi\frac{d^2u}{d\varphi^2}\right)=-f\cos\varphi.$$
Replacing in this equation $\dot{\varphi}$ by (\ref{eq31}), and simplifying we find equation (\ref{eq25}).

Equation (\ref{eq29}) implies that $\texttt{v}=\left|\underline{\dot{r}}\right|$ satisfies
\begin{equation}
\label{eq35}
\texttt{v}^2=\dot{r}^2+r^2\dot{\varphi}^2.
\end{equation}

Noting that
\begin{equation*}
\dot{r}=\frac{dr}{dt}=\dot{\varphi}\frac{d}{d\varphi}u^{-1}=-\dot{\varphi}u^{-2}\frac{du}{d\varphi},
\end{equation*}
and using for $\dot{\varphi}u^{-2}$ the expression obtained from (\ref{eq31}), namely
\begin{equation}
\label{eq36}
\dot{\varphi}u^{-2}=\frac{L}{m_r},
\end{equation}
it follows that
\begin{equation}
\label{eq37}
\dot{r}=-\frac{L}{m_r}\frac{du}{d\varphi}.
\end{equation}
Using in (\ref{eq35}) equations (\ref{eq36}) and (\ref{eq37}) we find eq. (\ref{eq26}).

\section{The perihelion precession of the Mercury}
In this case
\begin{equation}
\label{eq38}
m_r=m_o,\quad f=Gm_o\gamma^6\widetilde{u}^2.
\end{equation}
Thus, equations (\ref{eq25}) and (\ref{eq26}) become
\begin{equation}
\label{eq39}
\frac{d^2\widetilde{u}}{d\varphi^2}+\widetilde{u}=\frac{GM}{\widetilde{J}^2}\gamma^6,
\end{equation}
and
\begin{equation}
\label{eq40}
\widetilde{\texttt{v}}^2=\widetilde{J}^2\left[\widetilde{u}^2+\left(\frac{d\widetilde{u}}{d\varphi}\right)^2\right],
\end{equation}
where $\widetilde{J}$ denotes angular momentum per unit mass, i.e.
\begin{equation}
\label{eq41}
\widetilde{J}=\frac{\widetilde{L}}{m_o}.
\end{equation}
Introducing the dimensionless parameter $\widetilde{U}$ via
\begin{equation}
\label{eq42}
\widetilde{U}=\frac{\widetilde{J}^2}{GM}\widetilde{u}=\frac{\widetilde{p}}{\widetilde{r}},
\end{equation}
equations (\ref{eq39}) and (\ref{eq40}) become
\begin{equation}
\label{eq43}
\frac{d^2\widetilde{U}}{d\varphi^2}+\widetilde{U}=\gamma^6
\end{equation}
and
\begin{equation}
\label{eq44}
\frac{\widetilde{\texttt{v}}^2}{c^2}=\frac{\widetilde{\varepsilon}}{2}\left[\widetilde{U}^2+\left(\frac{d\widetilde{U}}{d\varphi}\right)^2\right],
\end{equation}
where $\widetilde{\varepsilon}$ is defined by
\begin{equation}
\label{eq45}
\widetilde{\varepsilon}=\frac{r_s}{\widetilde{p}}=2\left(\frac{GM}{c\widetilde{J}}\right)^2.
\end{equation}

Equations (\ref{eq43}) and (\ref{eq44}) together with the definition of $\gamma$ imply
\begin{equation}
\label{eq46}
\frac{d^2\widetilde{U}}{d\varphi^2}+\widetilde{U}=\left\{1-\frac{\widetilde{\varepsilon}}{2}\left[\widetilde{U}^2+\left(\frac{d\widetilde{U}}{d\varphi}\right)^2\right]\right\}^{-3}.
\end{equation}
In order to solve (\ref{eq46}) for small $\widetilde{\varepsilon}$ we let
\begin{equation}
\label{eq47}
\widetilde{U}=U_o+\frac{3}{2}\widetilde{\varepsilon}\widetilde{U}_1+O(\widetilde{\varepsilon}^2),\quad \quad \widetilde{\varepsilon}\rightarrow 0.
\end{equation}

Then, equation (\ref{eq46}) yields
\begin{equation}
\label{eq48}
\frac{d^2U_o}{d\varphi^2}+U_o=1,
\end{equation}
and 
\begin{equation}
\label{eq49}
\frac{d^2\widetilde{U}_1}{d\varphi^2}+\widetilde{U}_1=U^2_o+\left(\frac{dU_o}{d\varphi}\right)^2.
\end{equation}

The general solution of equation (\ref{eq48}) is
\begin{equation}
\label{eq50}
U_o=1+e\cos(\varphi-\varphi_o),
\end{equation}
where $e$ and $\varphi_o$ are constants. Letting for simplicity $\varphi_o=0$, equation (\ref{eq49}) becomes
\begin{equation}
\label{eq51}
\frac{d^2\widetilde{U}_1}{d\varphi^2}+\widetilde{U}_1=1+e^2+2e\cos\varphi .
\end{equation}
Thus,
\begin{equation}
\label{eq52}
\widetilde{U}_1=1+e^2+e\varphi\sin\varphi .
\end{equation}
Hence, using (\ref{eq50}) with $\varphi_o=0$ and (\ref{eq52}) into (\ref{eq47}) we find 
\begin{equation}
\label{eq53}
\widetilde{U}=1+e\cos\varphi+\frac{3}{2}\widetilde{\varepsilon}\left[1+e^2+e\varphi\sin\varphi\right].
\end{equation}
The dominant part of $\widetilde{U}$ denoted by $\widetilde{U}_D$ is given by
\begin{equation}
\label{eq54}
\widetilde{U}_D=\frac{3}{2}\widetilde{\varepsilon}e\varphi\sin\varphi.
\end{equation}
We next solve equation (\ref{eq7}) for small $\varepsilon$. Letting
\begin{equation}
\label{eq55}
U=U_o+\frac{3}{2}\varepsilon U_{1},
\end{equation}
we find that $U_{1}$ satisfies the ordinary differential equation
\begin{equation}
\label{eq56}
\frac{d^2U_{1}}{d\varphi^2}+U_{1}=1+e^2\cos^2\varphi+2e\cos\varphi.
\end{equation}
Thus,
\begin{equation}
\label{eq57}
U=1+e\cos\varphi+\frac{3}{2}\varepsilon\left[1+\frac{2}{3}e^2-\frac{e^2}{3}\cos^2\varphi+e\varphi\sin\varphi\right].
\end{equation}
The dominant part of $U$ denoted by $U_{D}$ is given by
\begin{equation}
\label{eq58}
U_{D}=\frac{3}{2}\varepsilon \varphi\sin\varphi.
\end{equation}

In order to relate $\varepsilon$ and $\widetilde{\varepsilon}$ we need to relate $J$ and $\widetilde{J}$. In this respect we note that equations (\ref{eq14}) and (\ref{eq24}) yield
\begin{equation}
\label{eq59}
J=r^2\frac{d\varphi}{d\tau},\quad \widetilde{J}=\widetilde{r}^2\frac{d\varphi}{dt}.
\end{equation}
Thus,
\begin{equation*}
\frac{J}{\widetilde{J}}=\left(\frac{\widetilde{u}}{u}\right)^2\frac{dt}{d\tau}.
\end{equation*}
Hence, using equation (\ref{eq13}) to compute $dt/d\tau$, as well as using the first of equations (\ref{eq8}) and equation (\ref{eq42}) to relate $u$ and $\widetilde{u}$ with $U$ and $\widetilde{U}$ we find the following:
\begin{equation*}
\frac{J}{\widetilde{J}}=\left(\frac{\widetilde{U}}{U}\right)^2\left(\frac{\widetilde{J}}{J}\right)^4\frac{E/mc^2}{1-\varepsilon U},
\end{equation*}
or
\begin{equation}
\label{eq60}
\left(\frac{J}{\widetilde{J}}\right)^3=\frac{E}{mc^2}\left(\frac{\widetilde{U}}{U}\right)^2\frac{1}{1-\varepsilon U}.
\end{equation}

Equation (\ref{eq18}) can be rewritten in the form
\begin{equation}
\label{eq61}
\left(\frac{dU}{d\varphi}\right)^2=\frac{2}{\varepsilon}\left(\frac{E}{mc^2}\right)^2-(1-\varepsilon U)\left(\frac{2}{\varepsilon}+U^2\right).
\end{equation}

Using in this equation the expansion 
\begin{equation*}
U=1+e\cos\varphi+O(\varepsilon),
\end{equation*}
it follows that
\begin{equation}
\label{eq62}
\frac{E}{mc^2}=1+\frac{\varepsilon}{4}(e^2-1)+O(\varepsilon^2),\quad \quad \varepsilon \rightarrow 0.
\end{equation}
Letting in (\ref{eq60})
\begin{equation*}
\frac{E}{mc^2}=1+O(\varepsilon),\quad \quad \varepsilon \rightarrow 0,
\end{equation*}
and
\begin{equation*}
\frac{\widetilde{U}}{U}=1+O(\varepsilon,\widetilde{\varepsilon}),\quad \varepsilon \rightarrow 0,\quad \widetilde{\varepsilon} \rightarrow 0,
\end{equation*}
we find
\begin{equation}
\label{eq63}
\frac{J}{\widetilde{J}}=1+O(\varepsilon,\widetilde{\varepsilon}),\quad \varepsilon \rightarrow 0,\quad \widetilde{\varepsilon} \rightarrow 0.
\end{equation}

Hence,
\begin{equation}
\label{eq64}
\widetilde{\varepsilon}=\varepsilon(1+O(\varepsilon)),
\end{equation}
thus $U_D$ coincides with $\widetilde{U}_D$.

The computation of the precession of the perihelion depends only on $U_D$. Indeed, equation (\ref{eq57}) implies that $dU/d\varphi$ vanishes at $\varphi=\varphi_*$, where $\varphi_*$ satisfies the equation 
\begin{equation}
\label{eq65}
\sin\varphi_*=\frac{3}{2}\varepsilon\left(-\frac{e}{3}\sin 2\varphi_*+\sin\varphi_*+\varphi_*\cos\varphi_*\right).
\end{equation}

\begin{figure}[t]
\vspace{1cm}
\begin{center}
\includegraphics[width=0.40\textwidth]{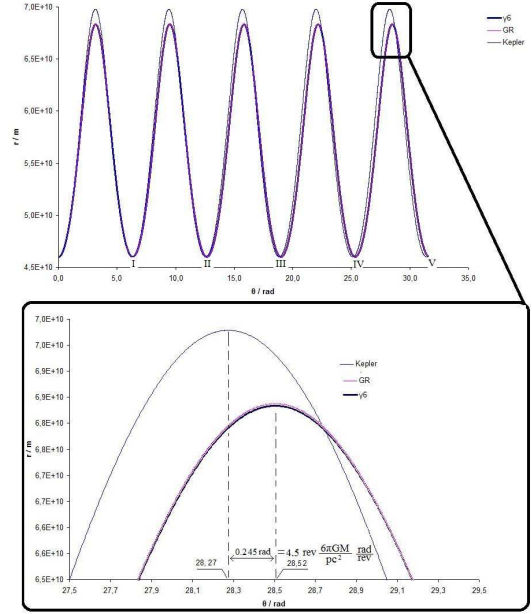}
\caption{Plots of the numerical integration of equations (\ref{eq7}) (general relativity, GR) and (\ref{eq9}) (relativistic gravitational law, RG) and comparison with the Keplerian orbit (\ref{eq48}) for reference. Distances in $10^{10}\;m$. For illustrative purposes the value used for GM is larger by a factor of $10^5$ than the real value of $1.3271 \times 10^{20}\; m^3/s^2$.}
\end{center}
\label{fig:2}
\end{figure}

One solution of this equation is $\varphi_*=0$. In order to obtain the second solution we let $\varphi_*=\pi+\varepsilon \omega$ and we use the identities
\begin{equation}
\label{eq66}
\cos(\pi+\varepsilon \omega)=-1+O(\varepsilon^2)\quad , \quad \sin(\pi+\varepsilon \omega)=-\varepsilon \omega+O(\varepsilon^3),\quad \varepsilon \rightarrow 0.
\end{equation}
Then, equation (\ref{eq65}) implies 
\begin{equation}
\label{eq67}
\varphi_*=\pi+\frac{3}{2}\pi\varepsilon=\pi+\frac{3}{2}\frac{\pi r_s}{p}=\pi+\frac{3\pi GM}{pc^2}.
\end{equation}
Equation (\ref{eq53}) with $\widetilde{\varepsilon}=\varepsilon$ also yields for $\varphi_*$ an expression identical with (\ref{eq66}).

Results of numerical integration of equations (\ref{eq7}) and (\ref{eq9}) as well as of the Kepler equation (\ref{eq48}) are presented in Figure 2, which shows that indeed the results of equations (\ref{eq7}) and (\ref{eq9}) coincide for small $\epsilon$, i.e. the $\gamma^6$ gravitational law yields to order $\epsilon$ the same results with general relativity.

\section {Conclusions}
The present results show that the relativistic Newton's gravitational Law, which uses gravitational rather than rest masses and thus contains the term $\gamma^3$ multiplying each moving mass in the familiar $Gm_1m_2/r^2$ or $GMm/r^2$ term, leads to a particle trajectory $r(\varphi)$ whose dominant term is the same with the dominant term of the familiar $r(\varphi)$ expression obtained via the theory of General Relativity. Furthermore, the perihelion precession angle, $\Delta\varphi$, computed via the former approach is equal to the value computed via General Relativity.

$$\textbf{Acknowledgements}$$
We thank Mr. D. Grigoriou for helpful discussions and for the numerical integration of  equations (7) and (9) shown in Figure 2.

\newpage
 \renewcommand{\theequation}{A-\arabic{equation}}
  \setcounter{equation}{0}
  \section*{APPENDIX A}
Here we establish the validity of the equation
\begin{equation}
\label{A1}
\texttt{v}\frac{d\texttt{v}}{dt}=\underline{\dot{r}}\cdot\underline{\ddot{r}}.
\end{equation}
Differentiating equation (\ref{eq29}) with respect to $t$ we find
\begin{equation}
\label{A2}
\underline{\ddot{r}}=\ddot{r}\underline{\hat{r}}+(2\dot{r}\dot{\varphi}+r\ddot{\varphi})\underline{\hat{r}}^+-r\dot{\varphi}^2\underline{\hat{r}}=(\ddot{r}-r\dot{\varphi}^2)\underline{\hat{r}}+(2\dot{r}\dot{\varphi}+r\ddot{\varphi})\underline{\hat{r}}^+,
\end{equation}
where the unit vector $\underline{\hat{r}}^+$ define by
\begin{equation}
\label{A3}
\underline{\hat{r}}^+=-\sin\varphi\underline{i}+\cos\varphi\underline{j},
\end{equation}
is orthogonal to the unit vector $\underline{\hat{r}}$.

Equations (\ref{eq28}), (\ref{eq29}), and (\ref{A3}) imply that $\underline{\dot{r}}$ can be rewritten in the form
\begin{equation}
\label{A4}
\underline{\dot{r}}=\dot{r}\underline{\hat{r}}+r\dot{\varphi}\underline{\hat{r}}^+.
\end{equation}

Hence using $\underline{\hat{r}}\cdot\underline{\hat{r}}^+=0$, it follows that
\begin{equation}
\label{A5}
\underline{\dot{r}}\cdot\underline{\ddot{r}}=(\ddot{r}-r\dot{\varphi}^2)\dot{r}+(2\dot{r}\dot{\varphi}+r\ddot{\varphi})r\dot{\varphi}.
\end{equation}

Thus,
\begin{equation}
\label{A6}
\underline{\dot{r}}\cdot\underline{\ddot{r}}=\dot{r}\ddot{r}+r\dot{r}\dot{\varphi}^2+r^2\dot{\varphi}\ddot{\varphi}.
\end{equation}
Differentiating (\ref{eq35}) with respect to $t$ we find
\begin{equation}
\label{A7}
\texttt{v}\frac{d\texttt{v}}{dt}=\dot{r}\ddot{r}+r\dot{r}\dot{\varphi}^2+r^2\dot{\varphi}\ddot{\varphi}.
\end{equation}

Equations (\ref{A6}) and (\ref{A7}) imply (\ref{A1}).

\newpage
 \renewcommand{\theequation}{B-\arabic{equation}}
  \setcounter{equation}{0}
  \section*{APPENDIX B}
We will establish the following fact: suppose that $r$ satisfies the equation
\begin{equation}
\label{B1}
\frac{d}{dt}(m_o\gamma \underline{\dot{r}})=-f\underline{\hat{r}}.
\end{equation}

Then, $H$ defined by
\begin{equation}
\label{B2}
H=m_o\gamma c^2+\texttt{V}(r),\quad f=\frac{d\texttt{V}}{dr},
\end{equation}
is constant. 

Indeed, equation (\ref{B2}) implies
\begin{equation}
\label{B3}
\frac{dH}{dt}=\left(m_oc^2\frac{d\gamma}{d\varphi}+\frac{d\texttt{V}}{dr}\frac{dr}{d\varphi}\right)\dot{\varphi}.
\end{equation}

Recall that
\begin{equation*}
\gamma^2=1+\frac{J^2}{c^2}\left[u^2+\left(\frac{du}{d\varphi}\right)^2\right].
\end{equation*}

Thus,
\begin{equation}
\label{B4}
\gamma\frac{d\gamma}{d\varphi}=\frac{J^2}{c^2}\left[u+\frac{d^2u}{d\varphi^2}\right]\frac{du}{d\varphi}.
\end{equation}

Furthermore, $u$ satisfies
\begin{equation*}
\frac{d^2u}{d\varphi^2}+u=\frac{\gamma}{m_oJ^2u^2}\frac{d\texttt{V}}{du}.
\end{equation*}

Hence, equation (\ref{B4}) becomes
\begin{equation*}
\frac{dH}{dt}=\left[\frac{m_oc^2}{\gamma}\frac{J^2}{c^2}\frac{\gamma}{m_oJ^2u^2}\frac{d\texttt{V}}{dr}+\frac{d\texttt{V}}{dr}\frac{d}{d\varphi}u^{-1}\right]\dot{\varphi}=0.
\end{equation*}
\end{document}